\documentclass[aps,twocolumn,amsmath,amssymb,showpacs]{revtex4}  
\usepackage{graphicx}  
\usepackage{dcolumn}   
\usepackage{bm}        
\usepackage{amssymb}   

\begin{document}

\title{Sharp knee phenomenon of primary cosmic ray energy spectrum}

\author{Samvel Ter-Antonyan}\altaffiliation{}\email{samvel_terantonyan@subr.edu}\affiliation{Department of Physics, Southern University, USA}
\date{\today}

\begin{abstract}
Primary energy spectral models are tested in the energy range of $1-200$ PeV using standardized 
extensive air shower responses from BASJE-MAS, Tibet, GAMMA and KASCADE
scintillation shower arrays. Results point towards the two-component origin of observed cosmic ray energy spectra 
in the knee region (GAPS spectral model) consisting of a pulsar component superimposed upon 
rigidity-dependent power law diffuse galactic flux. 
The two-component energy spectral model accounts for both the sharp knee shower spectral phenomenon and observed irregularity of all-particle
energy spectrum in the region of $50-100$ PeV.
Alternatively, tested multi-population primary energy spectra predicted by non-linear diffusive shock acceleration (DSA)
models describe observed shower spectra
in the knee region provided that the cutoff magnetic rigidities of accelerating particles are $6\pm0.3$ PV
and $45\pm2$ PV for the first two populations respectively.  Both tested spectral models confirm 
the predominant $H-He$ primary nuclei origin of observed shower spectral knee.
The parameters of tested energy spectra are evaluated using solutions of inverse
problem on the basis of the corresponding parameterizations of energy spectra
for primary $H$, $He$, $O$-like and $Fe$-like nuclei,
standardized shower size spectral responses in the $550-1085$ g/cm$^2$ 
atmospheric slant depth range and near vertical muon truncated size spectra
detected by the GAMMA array. 
\end{abstract}
\maketitle
\section{INTRODUCTION} 
The spectral knee phenomenon of primary cosmic ray energy spectrum in the region of $4-5$ PeV was discovered in 1958 \cite{Khrist}, while studying Extensive Air Showers (EAS) produced by high-energy primary nuclei in the atmosphere.   The change of the spectral power law index of detected EAS size spectrum pointed towards the corresponding change of primary energy spectral power index. The peculiarity of the knee phenomenon was not the change of spectral slope itself, but its high rate, which is still unresolved in the
frames of the standard models of the origin and propagation of galactic cosmic rays.

Until 1990s, the all-particle primary energy spectra derived from shower experiments were parameterized 
by a broken power law function  $F(E)\propto (E/E_k)^{-\gamma}$, where $\gamma\equiv\gamma_1\simeq2.7\pm0.03$ for $E<E_k$ and
$\gamma\equiv\gamma_2\simeq3.1\pm0.05$ for $E>E_k$ at knee energy $E_k\simeq3$ PeV.     
Appropriate approximation for the energy spectra of primary nuclei ($A\equiv H, He, ...Fe$) in the knee region taking
into account the rate of change of spectral slope was reported in \cite{samo2b}:   
\begin{equation}
F_A(E)=\Phi_AE^{-\gamma_1}
\left(1+\left(\frac{E}{E_k}\right)^\varepsilon\right)^\frac{\gamma_1-\gamma_2}{\varepsilon}\;,
\label{samo}
\end{equation}
where $E$ is the energy ($1<E<100$ PeV) of a primary nucleus $A$ with charge $Z$, $E_k=R\cdot Z$ is the  rigidity-dependent knee energy at which the asymptotic energy spectral power index $\gamma_1$ for $E\ll E_k$ is changed to the asymptotic value $\gamma_2$ for $E\gg E_k$ at sharpness parameter $\varepsilon$ correlating with the rate of change of the spectral slope. 

Expression (\ref{samo}) for sharpness parameter $\varepsilon=1$ can be derived from the superposition of energy spectra resulting from particle acceleration by the diffusive shock waves of Galactic supernova remnants (SNRs)  \cite{Bell,BV,nonlin} providing $S(E_c)\propto (E_c/E_k)^{\gamma_1-\gamma_2-1}\exp(-E_k/E_c)$ probability density function  \cite{Shibata} for
the maximal (cutoff) attainable energies ($E_c$)  in accelerating sites. However, the observed rate of change of the spectral slopes derived from EAS experiments in the knee region  \cite{EASTOP,KAS05,GAMAstro,Tibet,Chac,IceTop} actually corresponds to the energy spectral sharpness parameter $\varepsilon\gg1$  (so called "sharp knee" phenomenon).

Currently, the two phenomenological models of the origin and acceleration of Galactic cosmic rays can lay claim to the interpretation of this phenomenon: 1) the model describing the sharp knee origin by the contribution of nearby pulsar wind producing very hard particle energy spectra ($\sim E^{-1}$) \cite{Ostr,Blasi} to the power law diffuse Galactic cosmic ray flux in the knee region \cite{Bhadra,CERNC}; 2) the DSA spectral origin of the knee based on the theory of non-linear diffusive particle acceleration \cite{Bell,BV,nonlin} by shock waves driven by SNRs \cite{Hillas,Gaisser}. 
The common features of both models are the rigidity-dependent steepening of elemental ($A\equiv H,He,\dots Fe$) energy spectra \cite{Peters} and a multi-population spectral composition.

In this paper, the aforementioned two models of the origin of sharp knee phenomenon are tested using the parametrized solutions of inverse problem on the basis of standardized shower size spectra from  TIBET \cite{Tibet}, BASJE-MAS \cite{Chac}, KASCADE \cite{KAS05} and GAMMA \cite{GAMAstro} scintillation shower arrays. Primary energy spectra in the knee region obtained for each of the energy spectral models can be used to estimate the free parameters of corresponding theories \cite{Ostr,Bell,BV} for particle acceleration.

In Section~\ref{sec2} the main issues of the inverse problem of primary energy spectral unfolding are described. The standardization of shower size spectra  from different shower arrays \cite{Tibet,KASNe,IceTop,Chac,GAMJP} in $578-1085$ g/cm$^2$ atmospheric slant depth range are presented in Section~\ref{sec3}.  
The test of inverse problem solutions for different primary energy spectral models are presented in Section~\ref{sec4}. The interpretation of sharp spectral knee in terms of the pulsar wind contribution to the diffusive galactic cosmic ray flux (GAPS model) are discussed in Section~\ref{sec5}. 
\section{\label{sec2} INVERSE PROBLEM} 
The reconstruction of primary energy spectra $F_A(E)$ by the measured response $f({\bf{U}})$ of shower array (inverse problem, unfolding) is formulated via an integral equation
\begin{equation}
f({\bf{U}})=\sum_A\int {F_A(E)K_A(E,{\bf{U}})dE}\;,
\label{invprob}
\end{equation}
where $F_A(E)$  are object functions for primary nuclei $A$ with energy $E$ above the atmosphere and $K_A(E,{\bf{U}})$  are kernel functions describing  probability to detect and reconstruct air showers with a vector parameter ${\bf{U}}\equiv(N_e, N_{\mu}, \theta, ...)$. The sum in expression (\ref{invprob}) is calculated over all primary nuclei ($A\equiv H,He,...Fe$) or $N_A$ nuclei species ($H, He, CNO, Si$-like, $Fe$-like). 

Eq.~\ref{invprob} is a strongly ill-posed problem due to both, a set of object functions  and an $A$-dependence of the kernel function \cite{pseudo}. The theory of integral equations is not applicable to Eq.~\ref{invprob}. Even though the iteration unfolding algorithms for primary energy spectra \cite{KAS05,IceTop1,ICETOP} lead to plausible solutions,
the spectral errors of the solutions, as it is shown in \cite{pseudo}, are undetermined due to unavoidable inter-compensating pseudo solutions $F_A(E)+g_A(E)$
satisfying the condition
\begin{equation}
\sum_A\int {g_A(E)K_A(E,{\bf{U}})dE}=0(\pm\Delta f)
\label{pseudosol}
\end{equation}
for overall uncertainty of response function  $\Delta f(\bf{U})$  made of statistical errors and uncertainties of interaction model  \cite{pseudo}.

The unfolding of all-particle spectrum $F(E)=\sum_AF_A(E)$ from Eq.~\ref{invprob}  \cite{ICETOP,EASTOP}
requires {\em{a priori}} information about elemental energy spectra $F'_A(E)\simeq F_A(E)$
to compute the averaged kernel function $\overline{K(E)}=\sum F'_A(E)K_A(E)/\sum F'_A(E)$ over all primary nuclei $A$, which is an additional source of the systematic uncertainties 
of spectral solution $F(E)$. 

In the case of unfolding of the elemental primary energy spectra  $F_A(E)$ for $N_A>1$, the number of the possible combinations of pseudo solutions ($n_C$) satisfying condition (\ref{pseudosol}) 
increases rapidly with $N_A$ as $n_C=2^{N_A}-1$ \cite{pseudo}, which makes unfolding algorithms for Eq.~\ref{invprob} ineffective at  $N_A>3$. 
Examples of pseudo solutions for $N_A=4$ are shown in \cite{pseudo}.

 Pseudo solutions become apparent by varying the $N_{d.f.}$ initial (seed) values of iterative unfolding algorithms, where $N_{d.f.}\propto N_A$ is the number of the degrees of freedom for given $N_A$ object functions. On the other hand, the large number of object functions $N_A$ increasing the uncertainties of solutions will falsely improve the $\chi^2$ goodness-of-fit
test for expected and detected response functions, which is observed in \cite{KAS05} for $N_A=6$.\\

However, the inverse problem (\ref{invprob}) is transformed into the testing of parameterized primary energy spectra $F_A(E,|\Phi_A,\gamma_1,\gamma_2,\dots)$ like expression (\ref{samo})  or can be taken from a given model of the origin and acceleration of cosmic rays. Unknown spectral parameters  ($\Phi_A,\gamma_1,\gamma_2,\varepsilon,\dots$) can be estimated by the  $\chi^2$-test of measured shower spectra $f({\bf{U}})$ at the observation level by the expected response $f^*({\bf{U}})$ from the right hand side of Eq.~\ref{invprob} for the kernel function preliminary computed  in the frames of  a given interaction model. 

The advantage of the parametrized solutions of the inverse problem is not only in the lack of pseudo solutions but also in the reliable estimation of the errors of spectral parameters
provided that the number of spectral parameters is significantly lower than the number of the degrees of freedom for detected response $f({\bf{U}})$.

This approach, the so called parametrized regularization of the inverse problem, was implemented in \cite{BirGa} (${\bf{U}}\equiv N_e$) for AKENO \cite{AKENO} data, in \cite{SamBir} for MAKET-ANI data
(${\bf{U}}\equiv(N_{ch},\theta)$)  and in \cite{GAMAstro} for GAMMA array data (${\bf{U}}\equiv (N_{ch},N_{\mu},cos\theta,s)$). The application of this regularization method for GAMMA array data (${\bf{U}}\equiv (N_{ch},N_{\mu},cos\theta)$) \cite{GAMMA2013} and different primary spectral models are presented in Section~\ref{sec4}.
\section{\label{sec3}STANDARDIZED SHOWER SIZE SPECTRA} 
\begin{figure}
\includegraphics[scale=0.9]{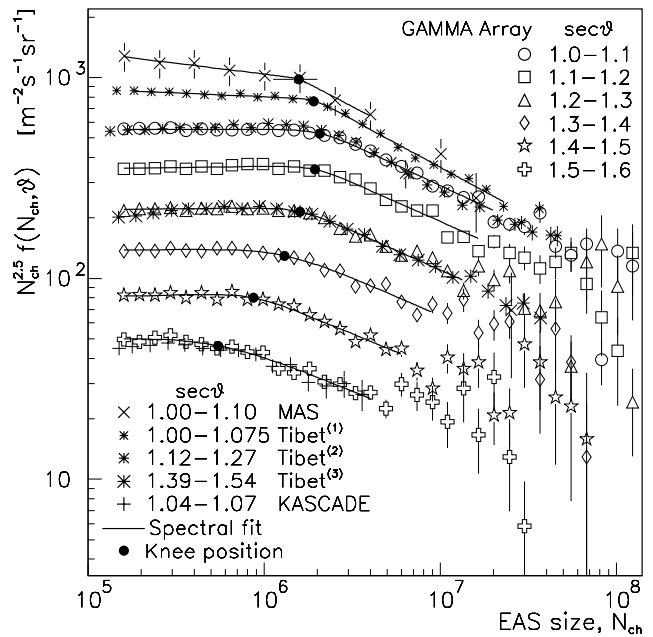}
\caption{\label{Nch} Standardized shower size spectra in the knee region for different atmospheric slant depths. Hollow symbols are GAMMA array data \cite{GAMAstro, GAM05}. The "$\times$", asterisk, and cross symbols are BASJE-MAS \cite{Chac}, standardized Tibet AS$\gamma$ \cite{Tibet} and KASCADE \cite{KASNe} shower size spectra correspondingly. Lines represent the approximations of shower size spectra according to expression (\ref{NchT}).  Solid circle symbols indicate the locations of corresponding spectral knees, $N_k(\theta)$. 
}
\end{figure}

To effectively solve the inverse problem (Eq.~\ref{invprob}) taking into account the sharpness of spectral knee, shower data from different experiments (observation levels) were studied using standardization of measurements. Detected responses $f({\bf{U}})$ at the knee region obtained from GAMMA experiment \cite{GAMJP,GAMAstro}  (observation level 700 g/cm$^2$) along with renormalized KASCADE \cite{KASNe} (1022 g/cm$^2$), Tibet AS$\gamma$ \cite{Tibet} (606 g/cm$^2$) and BASJE-MAS \cite{Chac} (550 g/cm$^2$) shower data are presented in Fig.~\ref{Nch} for different zenith angles $(\sec\theta)$. 

Shower size spectra from GAMMA array were considered in Fig.~\ref{Nch} as a standard, defining the detected shower size ($N_{ch}$) as the total number of shower charged particles with $E_e>1$ MeV energy threshold for electrons (positrons) \cite{GAMAstro}. The spectral data of KASCADE and Tibet AS$\gamma$ arrays in Fig.~\ref{Nch} were corrected (redefined) to the GAMMA array standard for $N_{ch}$ 
due to different definitions for the detected shower size ($N_{det}$) in the experiments \cite{GAMAstro,KASNe,Tibet,Chac}. Therefore the standardized spectral responses from different experiments in 
Fig.~\ref{Nch}  are homogeneous and can be used for spectral unfolding.   

Applied spectral correction,  $f(N_{ch})=\delta^{\gamma_N-1}f(N_{det})$, at a given correction factor (biases) for shower size $\delta=(1+N_{ch}/N_{det})$ stems from the log-normal distribution of biases, power law shower size spectra $f\simeq N_{ch}^{-\gamma_N}$ and a slight dependence of correction factor $\delta$ on the shower size $N_{ch}$ in the knee region \cite{GAMJP}.

The redefined KASCADE shower size spectrum ($N_{det}\equiv N_e$, \cite{KASNe}) in Fig.~\ref{Nch} takes into account the contribution of muon component $\delta_{\mu} =(1+N_{\mu}/N_e)$ and the energy threshold of detected electron component, $\delta_e = N_e(E_e>1MeV)/N_e(E_e>3MeV)$. Corrections $\delta_{\mu} =1.09\pm0.01$ and $\delta_{e}=1.15\pm0.01$ were computed using CORSIKA shower simulation code \cite{CORSIKA} for KASCADE observation level.

\begin{table}[b]
\caption{\label{tab0} Parameters of standardized shower size spectra from Fig.~1 (lines)
for different atmospheric slant depths.} 
\begin{ruledtabular}
\begin{tabular}{cccccc}
$\footnote{in the units of g/cm$^2$.}L$ &$\footnote{in the units of m$^{-2}\cdot$s$^{-1}\cdot$sr$^{-1}$.}\Phi_N$&$N_k/10^6$&$\varkappa$&$\gamma_{N,1}$&$\gamma_{N,2}$\\
\hline
578  &  4700$\pm$600 &1.6$\pm$0.7   &      $>$10        &  2.61$\pm$0.07& 2.96$\pm$0.08\\
629  &  1380$\pm$50  & 1.9$\pm$0.1   & 12$\pm$5      &  2.54$\pm$0.01& 2.97$\pm$0.02\\
735  &    552$\pm$4& 2.07$\pm$0.06   &  5.8$\pm$0.8 & 2.50$\pm$0.01& 2.91$\pm$0.03\\
805  &    314$\pm$3& 1.94$\pm$0.06 &  6.6$\pm$1.2 & 2.49$\pm$0.01& 2.90$\pm$0.03\\
875  &    195$\pm$2 & 1.59$\pm$0.05 &  5.6$\pm$1.1 & 2.49$\pm$0.01& 2.90$\pm$0.03\\
945  &    123$\pm$2 & 1.31$\pm$0.05 &  3.1$\pm$0.5 & 2.49$\pm$0.01& 2.89$\pm$0.04\\
1015&   64$\pm$1 & 0.86$\pm$0.03 &  4.9$\pm$1.5 & 2.48$\pm$0.02& 2.88$\pm$0.04\\
1085&   40$\pm$1 & 0.55$\pm$0.03 &  2.3$\pm$0.3 & 2.48$\pm$0.04& 2.88$\pm$0.05\\
\end{tabular}
\end{ruledtabular}
\end{table}

Standardized near-vertical Tibet$^{(1)}$ data in Fig.~1 have been computed using correction factors $\delta_{\gamma}^{(2)} =2.150\pm0.005$ 
and $\delta_{\gamma}^{(3)} =2.345\pm0.005$. Each correction factor was derived by the $\chi^2$-minimization of discrepancies between Tibet$^{(2,3)}$ data from \cite{Tibet}
and corresponding standard GAMMA shower size spectra  (Fig.~\ref{Nch}, hollow symbols) for the same atmospheric slant depths (Fig.~\ref{Nch}, two large asterisk symbols). The correction factor $\delta_{\gamma}^{(1)} =2.03\pm0.01$ for the near-vertical Tibet$^{(1)}$ spectrum in Fig.~\ref{Nch} 
(small asterisk symbols) was derived from the extrapolation of parameters $\delta_{\gamma}^{(2)}$ for $\overline{\sec{\theta}}\simeq1.2$ and $\delta_{\gamma}^{(3)}$  for $\overline{\sec{\theta}}\simeq1.47$   to the near-vertical Tibet spectrum at $\overline{\sec{\theta}}=1.038$. 

The dependence of correction factors $\delta_{\gamma}^{(1,2,3)}(\theta)$ on corresponding shower zenith angles ($\theta$) turned out to be in a close agreement with the expected
 attenuation of shower $\gamma$-quanta in the converter, $\delta_{\gamma}(\theta)-1\simeq(1-\exp{(-t\sec{\theta}/\lambda)}$, where $t=5.67$ g/cm$^2$ is
 the thickness of the lead converter [4] and $\lambda=15\pm2$ g/cm$^2$ is the attenuation length of shower  $\gamma$-quanta for average
 energy $E_{\gamma}\simeq30$ MeV.

The shower size spectrum of BASJE-MAS array in Fig.~\ref{Nch} was obtained unchanged ($\delta=1$) from the integral size spectrum \cite{Chac}  due to the identity of GAMMA and MAS scintillation detectors.

Lines in Fig.~\ref{Nch} are the approximations of shower size spectra in the knee region expressed by 
\begin{equation}
f(N_{ch})=
\Phi_N\cdot N_{ch}^{-\gamma_{N,1} }\left(1+\left(\frac{N_{ch}}{N_k}\right)^\varkappa\right)
^\frac{\Delta\gamma_N}{\varkappa}\;,
\label{NchT}
\end{equation}
where $\Delta\gamma_N=\gamma_{N,1}-\gamma_{N,2}$, and 
parameters  $\Phi_N$, spectral knee $N_k$ with sharpness $\varkappa$, 
 asymptotic spectral slopes $-\gamma_{N,1}$ and $-\gamma_{N,2}$ are presented in Table~I for different atmospheric slant depths $L=L_0\sec{\theta}$ at the $L_0$ location of shower array.

The key result stemming from Fig.~\ref{Nch} and Table~\ref{tab0} is the growth of shower spectral sharpness parameter from $\varkappa=2.3\pm0.3$ at $L=1085$ g/cm$^2$  to $\varkappa>6\pm0.5$  for high altitude measurements, where shower development is maximal ($dN_k(L)/dL\simeq0$) at minimal shower fluctuations.
 Because shower fluctuations described by the kernel function $K_A(E,N_{ch})$  from expression (\ref{invprob}) smooth away the sharpness of the shower spectral knee ($\varkappa$), the sharpness of the primary energy spectral knee ($\varepsilon$) should be at least more than the sharpness of shower spectral knee $\varkappa$. 
 
The evaluation of shower parameter $\varkappa$ from expressions (\ref{NchT}) at different energy spectral parameters $\varepsilon$ 
pointed towards relation
\begin{equation}
\varepsilon=\varkappa+(2\pm0.5)\gtrsim8.
\label{eps}
\end{equation} 
 The result (\ref{eps}) was obtained using the $\chi^2$-approximation of expected spectra $f^*({\bf{U}}\equiv N_{ch})$ from expressions (\ref{invprob}) and  (\ref{NchT})  
 at the log-normal kernel functions $K_A(E,N_{ch})$ and 
 primary energy spectra from  \cite{GAMAstro} for the $A\equiv H$ and $He$ nuclei responsible for shower spectral sharp knee at the observation level 700 g/cm$^2$.
\section{\label{sec4}TEST OF PARAMETERIZED SPECTRAL SOLUTIONS} 
\subsection{Kernel functions} 
The reconstructions of energy spectra in the knee region for $A\equiv H, He$ primary nuclei and $A\equiv O$-like and $Fe$-like nuclei species  were carried out on the basis of standardized shower spectra from Fig.~\ref{Nch} and near-vertical ($\sec\theta<1.2$) shower muon truncated ($r_{\mu}<100$ m) size spectra measured by the GAMMA array \cite{GAMJP,GAMMA2013} for 2003-2010.  The kernel functions $K_A(E|N_{ch},N_{\mu},\theta)$ for BASJE-MAS, Tibet, GAMMA and KASCADE  arrays were simulated by the CORSIKA code \cite{CORSIKA} in the frames of SIBYLL \cite{SIBYLL} interaction model for $A\equiv H,He,O$ and $Fe$ primary nuclei. Primary energies were simulated in the $0.5-500$ PeV region  using $F(E)\propto E^{-1.5}$ energy spectra providing approximately the same statistical errors in all energy regions.  

The kernel functions of all experiments were simulated obeying the GAMMA array standard \cite{GAMAstro,GAMJP} for the kinetic energy of shower particles: $E_e>1$ MeV, $E_\gamma>2$ MeV, $E_{\mu}>150$ MeV, $E_h>200$ MeV at the corresponding observation levels and geomagnetic fields.    
The right hand side of expression~(\ref{invprob}) was computed by the Monte-Carlo method.

\subsection{Sharp Knee spectral model} 
Sharp Knee phenomenological spectral model corresponds to the parameterization (\ref{samo}) for
sharpness parameter $\varepsilon=8$ from expression (\ref{eps}). 
\begin{table}[b]
\caption{\label{tab1} Scale parameters of Sharp Knee energy spectra (\ref{samo})
for $A\equiv H,He,O$ and $Fe$ primary nuclei at spectral parameters 
$\varepsilon=8\pm2$, $\gamma_1=2.68\pm0.015$, $\gamma_2=3.25\pm0.02$ and $E_k=R\cdot Z$ for $R=2900\pm200$ TV particle magnetic rigidity.}
\begin{ruledtabular}
\begin{tabular}{ccccc}
    $A$  &$H$&$He$&$O$&$Fe$ \\
\hline
\footnote{in the units of (m$^2\cdot$s$\cdot$sr$\cdot$TeV)$^{-1}$.}$\Phi_A$        &0.097$\pm$0.008&0.105$\pm$0.01&0.035$\pm$0.007&0.030$\pm$0.004      \\
\end{tabular}
\end{ruledtabular}
\end{table}
Spectral parameters $\Phi_A, \gamma_1,\gamma_2$ and $R$ were evaluated  from parametric Eq.~\ref{invprob} using the $\chi^2$-minimization 
of detected $f({\bf{U}})$ and expected $f^*({\bf{U}})$ spectral discrepancies. 
The regions of tolerances for spectral parameters were chosen to equal two standard errors ($2\sigma$) of corresponding values obtained  in the previous similar analysis 
of 2003-2007 GAMMA array data \cite{GAMAstro}. 
The evaluated parameters of Sharp Knee primary spectra (\ref{samo}) are presented in Table~\ref{tab1}.
 
\begin{figure}
\includegraphics[scale=0.9]{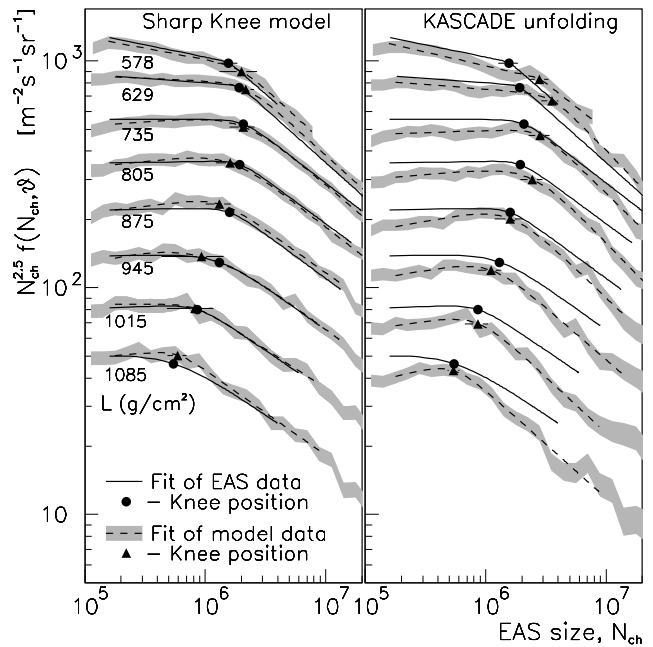}
\caption{\label{NchTak}
Standardized shower size spectral approximations (lines with solid circle symbol) replicated from Fig.~\ref{Nch} in comparison with corresponding expected size spectra 
(shaded areas) for different atmospheric slant depths $L(\theta)$ and different primary energy spectral models: Sharp Knee 
(expression (\ref{samo}), left panel) and KASCADE unfolded spectra from \cite{KAS05} (right panel). Dashed lines are the approximations of 
expected size spectra by the expression (\ref{NchT}) with corresponding shower spectral knee parameters, $N_k(\theta)$, (triangle symbols).  
}
\end{figure}

Expected shower size spectral responses $f^*(N_{ch},\theta)$ computed from the right hand side of expression (\ref{invprob}) are presented in Fig.~\ref{NchTak}  (left panel, shaded areas) in comparison with the corresponding approximations of standardized detected shower size spectra $f({\bf{U}})\equiv f(N_{ch},\theta)$ replicated from Fig.~\ref{Nch} (lines). 

The overall shower size spectrum ($\sec{\theta}<1.6$) and near-vertical ($\sec{\theta}<1.2$) shower muon truncated size spectrum 
obtained with GAMMA array in comparison with corresponding expected shower responses according to  Sharp Knee spectral model
are presented in Fig.~\ref{Ne_Nm} (hollow symbols). 

The obtained agreements of detected and expected shower size spectra correspond to $\chi^2_{\min}(\theta)\simeq1$ for energies up to about $50-70$ PeV for all atmospheric slant depths and describe the knee feature of shower spectra at the accuracies of less than 5\%.

\begin{figure}
\includegraphics[scale=0.9]{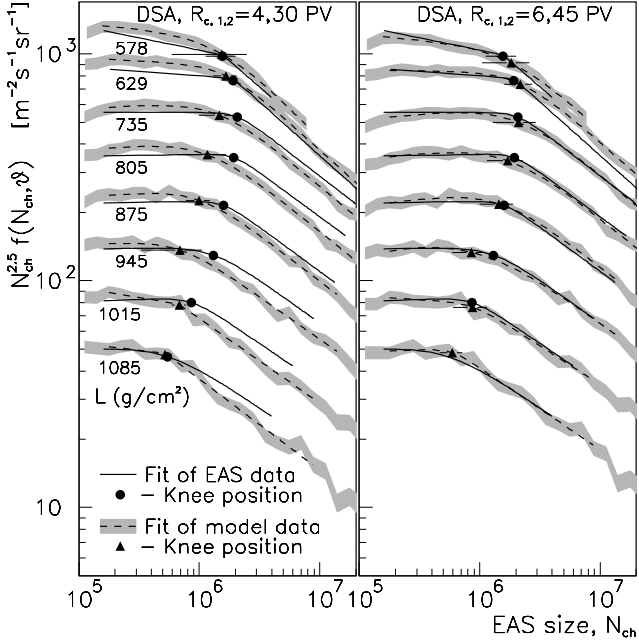}
\caption{\label{NchTgg} Same as Fig.~\ref{NchTak} for DSA energy spectral models: multi-population DSA model (\ref{gaiss}) from \cite{Gaisser}
(left panel, dashed lines) and DSA model with reevaluated cutoff particle magnetic rigidities, $R_{c,1,2}=6,\;45$ PV for the first two components respectively (right panel, dashed lines). 
}
\end{figure}

\subsection{KASCADE unfolded primary spectra} 
The expected shower size spectral responses $f^*(N_{ch},\theta$) computed from the right hand side of expression~(\ref{invprob}) for KASCADE unfolded primary 
spectra from \cite{KAS05}  are presented in the right panel of Fig.~\ref{NchTak} (shaded areas with dashed lines) in comparison with standardized shower size spectra $f(N_{ch},\theta$) (lines) replicated from Fig.~\ref{Nch}.

The overall response $f^*(N_{ch},\sec{\theta}<1.6)$ and shower muon response $f^*(N_{\mu},\sec{\theta}<1.2,r_{\mu}<100$ m$)$ 
expected from KASCADE unfolded energy spectra (dotted lines) \cite{KAS05}  in comparison with corresponding 
detected shower spectra obtained with GAMMA array (solid symbols) are presented in Fig.~\ref{Ne_Nm}.

The observed disagreements of expected  and detected shower data from Fig.~\ref{NchTak} (right panel) and  Fig.~\ref{Ne_Nm} 
can be explained by the $He-CNO$ nuclei origin of shower spectral knee resulting from the use of the Bayesian iterative unfolding algorithm in \cite{KAS05}  
which makes the primary composition in the knee region heavier (Section~\ref{sec2}, \cite{pseudo}) than it is expected from GAMMA array data \cite{GAMAstro}.

The common feature for both spectral predictions in Fig.~\ref{NchTak} (left and right panels) is the sharp spectral knees and the growth of knee sharpness with high altitude.
\subsection{Multi-population DSA spectral model} 
Expected shower spectral responses produced by multi-population DSA primary energy spectral model from \cite{Gaisser} 
\begin{equation}
F_{DSA}(E,A)=\sum_{i=1}^3\alpha_{A,i}E^{-\gamma_{A,i}}\exp\left(-\frac{E}{R_{c,i}Z}\right) 
\label{gaiss}
\end{equation}
were obtained from the right hand side of expression~(\ref{invprob}) at the cutoff magnetic rigidity of accelerated particles for the first two populations, $R_{c,1}=4$
 PV and  $R_{c,2}=30$ PV from  \cite{Gaisser}. The third ($i=3$), extragalactic population of energy spectra (\ref{gaiss}) can be ignored for the knee region.

The results of testing are presented in Fig.~\ref{NchTgg}, where the left panel shows the comparison of expected shower responses (shaded area with dashed lines ) with standardized
shower size spectra (solid lines ) replicated from Fig.~\ref{Nch}.

It is seen that despite a close agreement of expected and detected shower responses in the knee region, the detected shower sharp spectral knee feature 
is not reproduced and the expected shower spectral sharpness parameters are $\varkappa\simeq1.8-2.5$ for all atmospheric slant depths,
which is half the value observed in experiments (Table~\ref{tab0}).  
 \begin{figure}[t]
\includegraphics[scale=0.9]{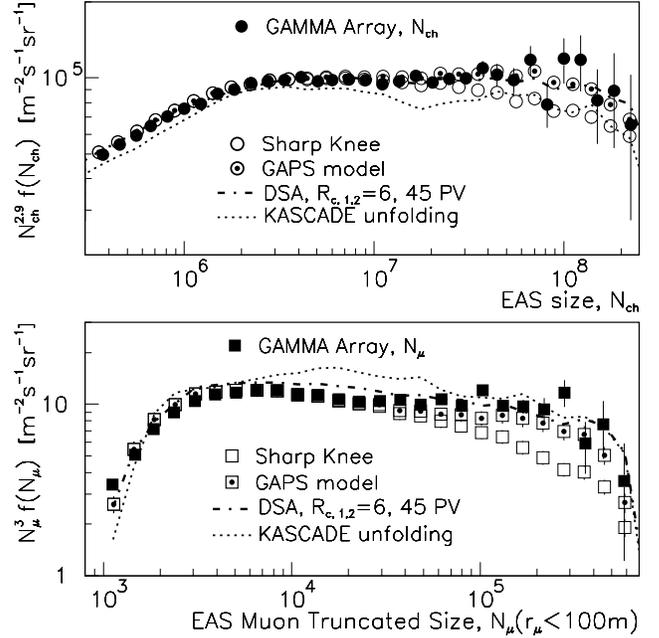}
\caption{\label{Ne_Nm}
Shower size spectrum, $f(N_{ch})$ (upper panel) and truncated muon size spectrum $f(N_{\mu}|E_{\mu}>4.6\cdot\sec\theta$ GeV$)$ (lower panel) from GAMMA array data \cite{GAMJP,PolCup}
(solid symbols) in comparison with corresponding different primary spectral predictions.}
\end{figure} 
To improve the agreement of expected and detected shower responses, the $R_{c,1,2}$ and $\alpha_{A,1,2}$ parameters of DSA spectral model from (\ref{gaiss}) were reevaluated
using the parameterized solution of Eq.~\ref{invprob} for standardized shower size spectra from Fig.~\ref{Nch} in the whole measurement range
and near-vertical shower muon truncated size spectra detected by the GAMMA array \cite{GAMJP,GAMMA2013}. 
The results are presented in Fig.~\ref{Ne_Nm}.

The observed agreement ($\chi^2\simeq 1$) was attained at the cutoff particle magnetic rigidities $R_{c,1}=6\pm0.3$ PV and  $R_{c,2}=45\pm2$ PV in
expression~(\ref{gaiss}). However, the expected shower spectral sharpness parameters turned out to be approximately the same, $\varkappa\lesssim2.5$.

The comparison of standardized detected shower size spectra with corresponding shower size spectral responses 
according to reevaluated DSA spectral model ($R_{c,1,2}=6$, $45$ PV) 
for different atmospheric slant depths are shown in Fig.~\ref{NchTgg} (right panel).

The same analysis for overall shower size spectrum ($\sec{\theta}<1.6$) and near-vertical ($\sec{\theta}<1.2$) shower muon truncated size spectrum
are presented in the upper and lower panels of Fig.~\ref{Ne_Nm} (dash-dotted lines) correspondingly.

It is seen that reevaluated DSA spectral model describes the detected shower responses in full measurement ranges including
irregularities in the energy region of $50-100$ PeV.
The reevaluated values of scale parameters $\alpha_{A,1,2}$ from expression~(\ref{gaiss}) are presented in Table~\ref{tabg}.
\begin{table}[b]
\caption{\label{tabg} Reevaluated DSA primary energy spectral scale parameters $\alpha_{A,1,2}$ 
in comparison with original values from \cite{Gaisser} for different primary nuclei $A$.
}
\begin{ruledtabular}
\begin{tabular}{ccccc}
$A$ & $\footnote{in the units of $(m^2\cdot s\cdot sr\cdot GeV)^{-1}$.} \alpha_{A,1}$ &$\alpha_{A,1}$\cite{Gaisser} &$\alpha_{A,2}$ &$\alpha_{A,2}$ \cite{Gaisser}\\
\hline
 p        &  $7500\pm610$ &7860&$25\pm7$ &20\\
 He     &  $3000\pm290$ &3550&$20\pm4$ &20\\
 CNO  &  $1500\pm300$ &2200&$10\pm3$ &13.4\\
 Mg-Si &  $500\pm150$   &1450& $7\pm5$  &13.4\\
  Fe      &  $2120\pm250$ &2120& $13.4\pm3$ &13.4\\
\end{tabular}
\end{ruledtabular}
\end{table}
\subsection{GAPS spectral model} 
The observed GAMMA array shower spectral irregularities  in the region of $E>50-100$ PeV  (Fig.~\ref{Nch}) are not
described by expression (\ref{samo}) by definition and indicate the occurrence of an additional $Fe$ component 
with energy spectrum $\propto E^{-1\pm0.5}$ \cite{GAMJP}.  The model of particle acceleration 
by the pulsar wind can provide such a hard energy spectrum ($\sim E^{-1}$) {\cite{Ostr,Blasi}.
 
Here, the concept of two-component $Fe$ flux in the region of $70-100$ PeV from \cite{GAMJP} was tested
for all primary nuclei to describe the sharp knee phenomenon. 
Two-component energy spectra for $A\equiv H,He,O$ and $Fe$ primary nuclei in the knee region were parameterized by the expression
\begin{equation}
F_A(E)=F_G(A,E)+F_P(A,E)
\label{twocomp}
\end{equation}
composed of the diffuse galactic cosmic ray flux  $F_G(A,E|\varepsilon=1)$
from expression (\ref{samo})
and a particle flux accelerated by pulsar wind \cite{Ostr,Blasi}
\begin{equation}
F_P(A,E)=\Psi_{A}E^{-(1+\eta)}e^{-\frac{E}{E_c}}
\label{pulsar}
\end{equation}
 taking into account the leakage of particles from a confinement volume (local Superbubble\cite{bubble}) with rate $\propto E^{-\eta}$.
Hereinafter the primary energy spectral model from expressions (\ref{twocomp},\ref{pulsar})  is called GAPS ({\bf{G}}alactic {\bf{A}}nd {\bf{P}}ulsar {\bf{S}}uperposition) model. 
\begin{figure}
\includegraphics[scale=0.9]{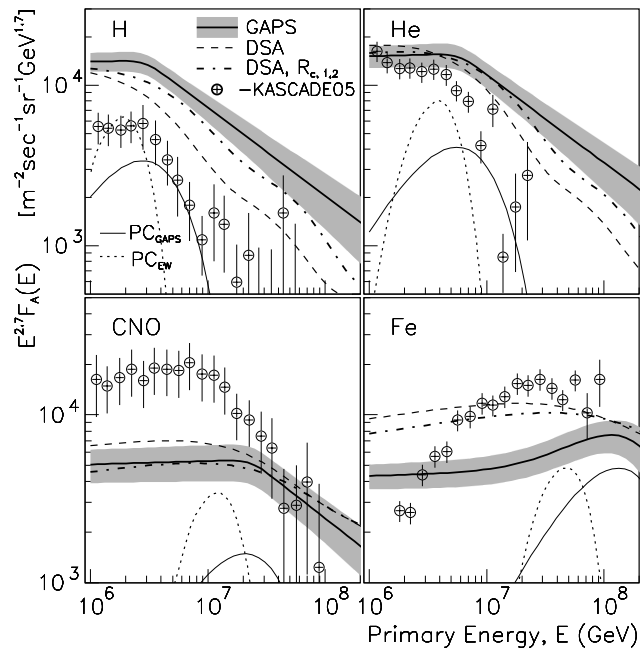}
\caption{\label{knee_e}Energy spectra according to GAPS, DSA \cite{Gaisser} and reevaluated DSA (Table~\ref{tabg}) spectral models
for primary $H$, $He$, $O$-like and $Fe$-like nuclei.  The symbols are KASCADE unfolded primary energy spectra from \cite{KAS05}. 
The thin solid (PC$_{GAPS}$) and dotted (PC$_{EW}$) lines are the corresponding expected energy spectra of pulsar components from (\ref{pulsar}) and
\cite{CERNC} respectively.
}
\end{figure}

The scale parameters $\Psi_A$ and the maximal (cutoff) energy $E_c(A)$ of particles accelerated by a pulsar wind
in expression (\ref{pulsar}) are estimated by solving Eq.~\ref{invprob} on the basis of
GAMMA array data and parameterizations (\ref{samo},\ref{twocomp},\ref{pulsar}) (Section~\ref{sec5}).

The results of the overall shower size spectrum and the truncated muon size spectrum  of GAMMA
array \cite{GAMJP,GAMMA2013} are presented in Fig.~\ref{Ne_Nm} along with expected responses according to the Sharp Knee energy spectra  
(hollow symbols).  The expected shower spectral responses corresponding to the GAPS primary spectral model from expressions (\ref{twocomp},\ref{pulsar})  
 are shown by the circle dot symbols (upper panel) and square dot symbols (lower panel).
 Dashed-dot lines and dashed lines in Fig.~\ref{Ne_Nm} are the expected responses obtained from KASCADE \cite{KAS05} and reevaluated DSA \cite{Gaisser} energy spectra. 
 
 Good agreement between detected and expected shower responses is noted for both GAPS and reevaluated DSA 
primary energy spectral models.\\

The review of the parameterized solutions of Eq.~\ref{invprob} for the energy spectra of $A\equiv H,He,O (CNO)$ and $Fe$ primary nuclei in the energy range of $1-200$ PeV
(lines) and KASCADE unfolded energy spectra (symbols) are presented in Fig.~\ref{knee_e}.
 
\begin{figure}
\includegraphics[scale=0.9]{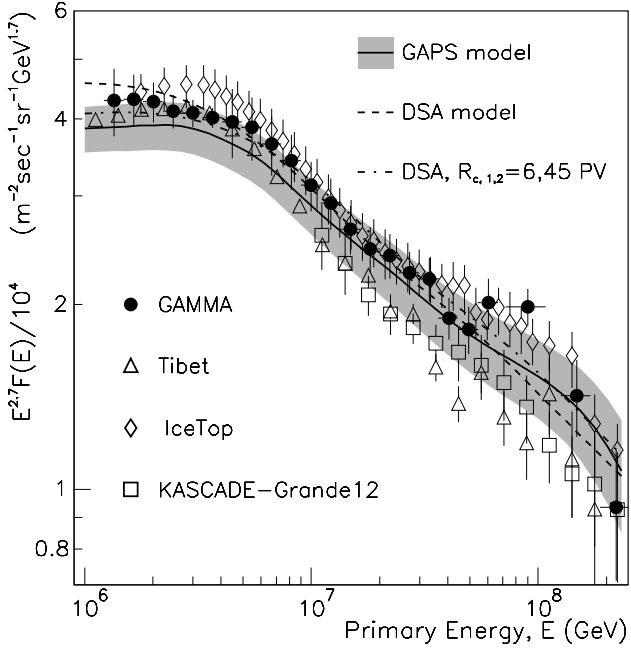}
\caption{\label{knee_all} All-particle energy spectra from GAPS  (line with shaded area), DSA  \cite{Gaisser} and reevaluated
DSA spectral models (dash-dotted line). The symbols represent the experimental data from \cite{GAMJP,Tibet,IceTop,GRANDE}
obtained using event-by-event primary energy evaluations.
}
\end{figure}
\begin{table}
\caption{\label{tab3} All-particle primary energy spectrum
in the units of $(m^2\cdot s\cdot sr\cdot GeV)^{-1}$ obtained with GAMMA shower array.} 
\begin{ruledtabular}
\begin{tabular}{cccc}
$E/PeV$ &$dF/dE\pm\Delta_{tot}$&$E/PeV$ &$dF/dE\pm\Delta_{tot}$\\
\hline
 1.35 &  (1.20$\pm$0.15)$\cdot10^{-12}$&14.9 &  (1.14$\pm$0.10)$\cdot10^{-15}$\\
  1.65 &  (7.04$\pm$0.67)$\cdot10^{-13}$&18.2 &  (6.15$\pm$0.53)$\cdot10^{-16}$\\
  2.01 &  (4.09$\pm$0.31)$\cdot10^{-13}$&22.2 &  (3.51$\pm$0.29)$\cdot10^{-16}$\\
  2.46 &  (2.29$\pm$0.14)$\cdot10^{-13}$&27.1 &  (1.92$\pm$0.15)$\cdot10^{-16}$\\
  3.00 &  (1.33$\pm$0.07)$\cdot10^{-13}$&33.1 &  (1.09$\pm$0.09)$\cdot10^{-16}$\\
  3.67 &  (7.58$\pm$0.28)$\cdot10^{-14}$&40.4 &  (5.51$\pm$0.44)$\cdot10^{-17}$\\
  4.48 &  (4.36$\pm$0.56)$\cdot10^{-14}$&49.4 &  (3.07$\pm$0.27)$\cdot10^{-17}$\\
  5.47 &  (2.49$\pm$0.30)$\cdot10^{-14}$&60.3 &  (1.98$\pm$0.19)$\cdot10^{-17}$\\
  6.69 &  (1.36$\pm$0.15)$\cdot10^{-14}$&90.0 &  (6.61$\pm$0.45)$\cdot10^{-18}$\\
  8.17 &  (7.43$\pm$0.76)$\cdot10^{-15}$&148 &  (1.24$\pm$0.18)$\cdot10^{-18}$\\
  9.97 &  (3.96$\pm$0.37)$\cdot10^{-15}$&221 &  (2.76$\pm$0.66)$\cdot10^{-19}$\\
 12.2 &  (2.15$\pm$0.21)$\cdot10^{-15}$&- &-\\
\end{tabular}
\end{ruledtabular}
\end{table}

Corresponding expected all-particle energy spectra for aforementioned spectral models are shown in Fig.~\ref{knee_all} 
in comparison with measurements (symbols) using event-by-event primary energy reconstructions from \cite{GAMMA2013,Tibet,IceTop,GRANDE} shower arrays.
The all-particle spectrum obtained with GAMMA array \cite{GAMMA2013} is presented in Table~\ref{tab3}.
\section{\label{sec5}Sharp knee and GAPS spectral models} 
\begin{table}[b]
\caption{\label{tab4} Parameters of the energy spectra of pulsar component (\ref{pulsar}) for $A\equiv H,He,O,Fe$ nuclei and $\eta=0.35$.}
\begin{ruledtabular}
\begin{tabular}{ccccc}
    A   &$H$&$He$&$O$&$Fe$ \\
\hline
\footnote{in the units of (m$^2\cdot$s$\cdot$sr$\cdot$TeV)$^{-1}$.}
$\Psi_A$        &2.3$^{+.2}_{-.5}\times$10$^{-6}$ &1.1$^{+.1}_{-.1}\times$10$^{-6}$ & 6.7$^{+.9}_{-.8}\times$10$^{-8}$&1.8$^{+.2}_{-.2}\times$10$^{-8}$      \\
$E_c/E_k$    &            0.71$\pm$0.06               &           0.71$\pm$0.04                  &           0.67$\pm$0.06                &              1.30$\pm$0.08                   \\                                          
\end{tabular}
\end{ruledtabular}
\end{table}
\begin{table*}[t]
\caption{\label{tab5}Energy spectra ($E_A^{2.7}\cdot dF/dE_A$) for $A\equiv H, He,O$ and $Fe$ primary nuclei
in the units of $m^{-2}\cdot s^{-1}\cdot sr^{-1}\cdot TeV^{1.7}$ according to GAPS spectral model. Spectral uncertainties include statistical and systematic errors.} 
\begin{ruledtabular}
\begin{tabular}{cccccccccc}
$E_A$/PeV&$H$&$He$&$O$&$Fe$&$E_A$/PeV&$H$&$He$&$O$&$Fe$\\
\hline
 1.00 & 0.111$\pm$0.015 & 0.121$\pm$0.019 & 0.040$\pm$0.009 & 0.034$\pm$0.006&15.41 & 0.045$\pm$0.010 & 0.073$\pm$0.015 & 0.042$\pm$0.010 & 0.040$\pm$0.008\\
  1.20 & 0.112$\pm$0.015 & 0.121$\pm$0.019 & 0.040$\pm$0.009 & 0.035$\pm$0.006&18.49 & 0.041$\pm$0.010 & 0.066$\pm$0.014 & 0.042$\pm$0.010 & 0.041$\pm$0.008\\
  1.44 & 0.112$\pm$0.015 & 0.121$\pm$0.019 & 0.041$\pm$0.009 & 0.035$\pm$0.006&22.19 & 0.037$\pm$0.009 & 0.060$\pm$0.013 & 0.042$\pm$0.010 & 0.042$\pm$0.008\\
  1.73 & 0.113$\pm$0.016 & 0.122$\pm$0.019 & 0.041$\pm$0.009 & 0.035$\pm$0.006&26.62 & 0.033$\pm$0.009 & 0.054$\pm$0.012 & 0.039$\pm$0.010 & 0.044$\pm$0.009\\
  2.07 & 0.113$\pm$0.016 & 0.122$\pm$0.020 & 0.041$\pm$0.009 & 0.035$\pm$0.006&31.95 & 0.030$\pm$0.009 & 0.049$\pm$0.012 & 0.036$\pm$0.009 & 0.046$\pm$0.009\\
  2.49 & 0.111$\pm$0.016 & 0.123$\pm$0.020 & 0.041$\pm$0.010 & 0.035$\pm$0.006&38.34 & 0.028$\pm$0.008 & 0.044$\pm$0.012 & 0.032$\pm$0.008 & 0.048$\pm$0.010\\
  2.99 & 0.107$\pm$0.017 & 0.123$\pm$0.020 & 0.041$\pm$0.010 & 0.035$\pm$0.006&46.01 & 0.025$\pm$0.008 & 0.040$\pm$0.011 & 0.029$\pm$0.008 & 0.051$\pm$0.011\\
  3.58 & 0.100$\pm$0.015 & 0.124$\pm$0.021 & 0.041$\pm$0.010 & 0.036$\pm$0.006&55.21 & 0.023$\pm$0.007 & 0.036$\pm$0.010 & 0.027$\pm$0.007 & 0.053$\pm$0.011\\
  4.30 & 0.091$\pm$0.014 & 0.123$\pm$0.021 & 0.041$\pm$0.010 & 0.036$\pm$0.007&66.25 & 0.020$\pm$0.007 & 0.033$\pm$0.010 & 0.024$\pm$0.007 & 0.056$\pm$0.012\\
  5.16 & 0.083$\pm$0.013 & 0.122$\pm$0.021 & 0.042$\pm$0.010 & 0.036$\pm$0.007& 79.50 & 0.018$\pm$0.007 & 0.030$\pm$0.009 & 0.022$\pm$0.006 & 0.058$\pm$0.012\\
  6.19 & 0.075$\pm$0.012 & 0.117$\pm$0.020 & 0.042$\pm$0.010 & 0.036$\pm$0.007& 95.40 & 0.017$\pm$0.006 & 0.027$\pm$0.009 & 0.020$\pm$0.006 & 0.060$\pm$0.013\\
  7.43 & 0.068$\pm$0.012 & 0.108$\pm$0.019 & 0.042$\pm$0.011 & 0.037$\pm$0.007&114.5 & 0.015$\pm$0.006 & 0.024$\pm$0.008 & 0.018$\pm$0.005 & 0.060$\pm$0.013\\
  8.92 & 0.061$\pm$0.011 & 0.100$\pm$0.018 & 0.042$\pm$0.010 & 0.037$\pm$0.007&137.4 & 0.014$\pm$0.006 & 0.022$\pm$0.008 & 0.016$\pm$0.005 & 0.059$\pm$0.014\\
 10.70 & 0.055$\pm$0.011 & 0.089$\pm$0.016 & 0.042$\pm$0.010 & 0.038$\pm$0.007&164.8 & 0.012$\pm$0.005 & 0.020$\pm$0.007 & 0.015$\pm$0.005 & 0.057$\pm$0.015\\
 12.84 & 0.050$\pm$0.010 & 0.081$\pm$0.015 & 0.042$\pm$0.010 & 0.039$\pm$0.008&197.8 & 0.011$\pm$0.005 & 0.018$\pm$0.007 & 0.013$\pm$0.004 & 0.052$\pm$0.015\\
\end{tabular}
\end{ruledtabular}
\end{table*}

Applying the two-component origin of energy spectra in the knee region (\ref{twocomp},\ref{pulsar}) to all nuclei species, the sharp knee spectral phenomenon
can be interpreted in the frames of the GAPS spectral model. Results are presented in Fig.~\ref{knee_pc}.
\begin{figure}[t]
\includegraphics[scale=0.9]{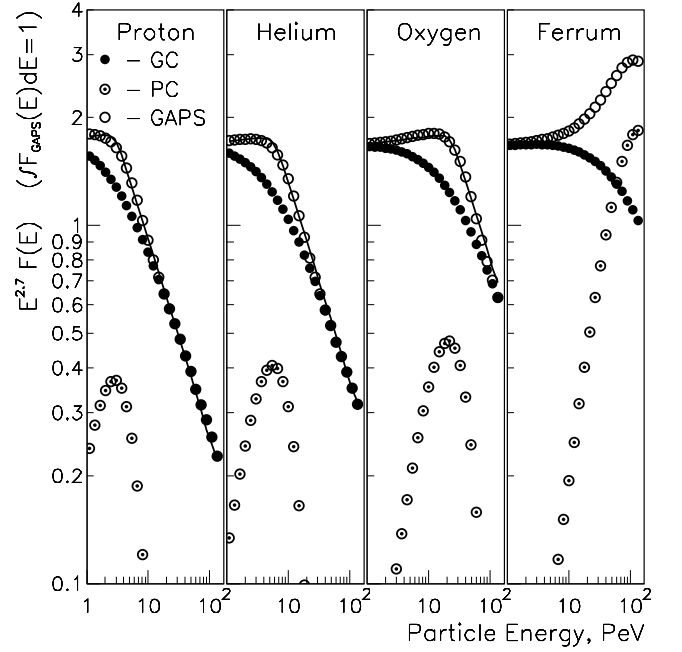}
\caption{\label{knee_pc} Normalized energy spectra for $H,He,O$ and $Fe$ primary nuclei
approximated by expression  (\ref{twocomp}) according to the GASP spectral model composed of the pulsar component (PC) from expression (\ref{pulsar})   
and the diffuse galactic component (GC) from expression (\ref{samo}) at $\varepsilon=1$. Lines are the corresponding 
energy spectra according to the Sharp Knee spectral model for $H,He$ and O-like nuclei from expression (\ref{samo}) at $\varepsilon=8$.
}
\end{figure}
The spectral parameters $\Psi_A$ and $E_c(A)$ of energy
spectra for the $H,He,O$ and $Fe$ primary nuclei from the pulsar wind (\ref{pulsar}) are presented in Table~\ref{tab4}. 
The parameters of diffuse galactic component are the same as the parameters of sharp knee spectra
(expression (\ref{samo}) and Table~\ref{tab1}) except for parameter $\varepsilon=1$.

The obtained energy spectra of pulsar components according to the GAPS spectral model are presented in Fig.~\ref{knee_e} (thin solid lines)
in comparison with corresponding estimations from \cite{CERNC} (dotted lines).   

The evaluated values of spectral parameters $E_c(A)$ from Table~\ref{tab4} 
for $H-O$ nuclei turned out to be rigidity dependent  
whereas the maximal energy of iron pulsar component, 
$E_c(Fe)\simeq100$ PeV, is about twice as high as it should be. The obtained large magnetic rigidity for the iron nuclei
of pulsar component could be an indication of the presence of a second younger ($<10^4$ years) pulsar in the same confinement volume,
though a possible contribution of extragalactic population \cite{Gaisser} in the energy range $E\simeq100-200$ PeV can no longer be excluded.\\

Existing skepticism about the low efficiency of particle acceleration by pulsars is mainly associated 
with the high cooling rate of pulsars and the corresponding low efficiency of thermionic emission from the surface into the magnetosphere of a pulsar.
 In this respect, particle eruptions into the magnetosphere due to a possible volcanic activity 
 of pulsars proposed in \cite{volcano,volcano2} could provide the required particle density in the magnetosphere.
 
 Assuming dynamic equilibrium between volcanic material erupting onto the magnetosphere of a pulsar and particle flux accelerated 
 by the pulsar wind, the  
 confinement volume for pulsar component can be estimated from particle flux-density relationship \cite{Gaisserbook}, 
 \[
 \Im=\frac{\rho_p\beta c}{4\pi}\;, 
 \]
where
$c$ is the speed of light, $\beta\simeq1$ is a particle speed, $\Im=\int{F_P(E)dE}$ is a detected particle flux in the units of $cm^{-2}\cdot s^{-1}\cdot sr^{-1}$ 
and $\rho_p=N_{p,tot}/V_c$ is a particle density in a confinement volume $V_c$. 

The predicted rate of eruption material $M>10^6$ g$\cdot$cm$^{-2}\cdot$s$^{-1}$ from \cite{volcano}, the integral spectrum of the pulsar 
 proton component from (\ref{pulsar}) and Table~\ref{tab4}, $\Im(E_p>10$ GeV$)=3.2\cdot10^{-9}$ cm$^{-2}\cdot$s$^{-1}\cdot$sr$^{-1}$
along with suggested permanent eruption time $t=10^4$ years from the total of $A=10^3$ cm$^2$ erupted surface area of a pulsar 
 result in confinement volume 
 \begin{equation}
 V_c=\frac{MN_ActA}{4\pi\Im}\simeq1.4\cdot10^{62}\; cm^3\;,
 \label{confV}
 \end{equation}
 where $N_A$ is Avogadro number. The corresponding radius of confinement volume  is $r_c\gtrsim100$ pc, which is well in agreement  with the size of the local Superbubble \cite{bubble}.

The average energy of the pulsar component from (\ref{pulsar}), $\overline{E_p}\simeq14$ TeV,
determines the upper limit for the corresponding energy density of the pulsar component in the cavity of the Superbubble
$\rho_E=\rho_{p}\overline{E_p}\simeq2\cdot10^{-5}$ eV/cm$^3$. This value is 
negligible compared to the galactic cosmic ray energy density, $\sim1$ eV/cm$^3$, albeit is enough for the formation of the sharp spectral 
knee phenomenon.


\section{Summary}
The standardization of shower spectral responses turned out to be an effective tool
for testing of the primary energy spectral models.

Two phenomenological energy spectral models have been tested 
using the parameterized solution of the inverse problem by the $\chi^2$-minimization of the discrepancies of expected and detected shower responses
in a broad atmosphere slant depth range ($550-1085$ g/cm$^2$) for primary $H$, $He$, $O$-like and $Fe$-like nuclei in the energy range $1-200$ PeV.

The GAPS spectral model (expression (\ref{twocomp})) formed from a pulsar component (\ref{pulsar}) superimposed upon the
rigidity-dependent steepening power law diffuse galactic flux (expression (\ref{samo}) for $\varepsilon=1$) describes both the shower responses and the dependence of shower sharpness
parameters $\varkappa$ on atmosphere slant depths (Table~\ref{tab0}). This result confirms the local origin of the sharp knee phenomenon from \cite{Bhadra,CERNC}.
Energy spectra according to the GAPS spectral model from Fig.~\ref{knee_e} are presented in Table~\ref{tab5}.
 
The multi-population DSA spectral model from expression~(\ref{gaiss}) can describe 
observed shower responses provided that spectral cutoff particle magnetic rigidities are $R_{c,1}=6.0\pm0.3$ PV and $R_{c,2}=45\pm2$ PV for the first two spectral populations (Table~\ref{tabg})
which is 1.5 times greater than it is predicted in \cite{Gaisser}. However, the observed shower spectral sharpness parameter from expression (\ref{NchT}) is not reproduced by the DSA spectral model for high altitudes ($\varkappa\simeq6$, Table~\ref{tab0}) and remains approximately constant at about $\varkappa\lesssim2.5$ for all atmospheric slant depths.  

Both spectral models confirm the predominant $H-He$ origin of the observed shower spectral knee and can describe the flattening \cite{GRANDE} of the all-particle energy spectrum 
in the range of $50-100$ PeV (Fig.~\ref{knee_all}). 

The obtained phenomenological pulsar wind component can be produced by the mature Geminga pulsar (age $\sim3\cdot10^5$ years, distance $\sim250$ pc) \cite{Bhadra}
being  the possible cause of the local Superbubble \cite{source}, provided that the hypothesis of the volcanic activity is confirmed.

\end{document}